# Dynamic control of the optical emission from GaN/InGaN nanowire quantum dots by surface acoustic waves

S. Lazić,[1,a)], E. Chernysheva,[1] Ž. Gačević,[2] H.P. van der Meulen,[1] E. Calleja[2] and J.M. Calleja Pardo[1]

[1]*Departamento de Física de Materiales, Instituto "Nicolás Cabrera" and Instituto de Física de Materia Condensada (IFIMAC), Universidad Autónoma de Madrid, 28049 Madrid, Spain*

[2]*ISOM-DIE, Universidad Politécnica de Madrid, 28040 Madrid, Spain*

The optical emission of InGaN quantum dots embedded in GaN nanowires is dynamically controlled by a surface acoustic wave (SAW). The emission energy of both the exciton and biexciton lines is modulated over a 1.5 meV range at ~330 MHz. A small but systematic difference in the exciton and biexciton spectral modulation reveals a linear change of the biexciton binding energy with the SAW amplitude. The present results are relevant for the dynamic control of individual single photon emitters based on nitride semiconductors.

Surface acoustic waves induce periodic strain and piezoelectric fields near a semiconductor surface which can dynamically modify their basic properties. The use of SAWs is an expanding research field, which has been widely applied to semiconductor quantum wells (QWs)[1-3], wires[4-6] and dots (QDs)[7-11].

By controlling the excitonic emission in III-V semiconductor QDs by SAWs, high repetition rate single photon sources (SPSs)[7-9] and periodic laser mode feeding[12] have been reported. Also, dynamic control of individual QDs[11] and on-demand single-electron transfer between distant quantum dots[13] have been demonstrated. In a more general context, a proposal for on-chip quantum transducers based on SAWs enabling long-range coupling of many qubits has been recently put forward[14]. In addition to the band-edge modulation, which determines the QD emission wavelength, the tunable strain field of the SAW can be used to modify other properties related to the band structure, as the exciton binding energy, in a similar way as static strain field[15].

While most of the SAW-related work reported in semiconductor structures refers to III-V systems, studies on group III-Nitrides are scarce. Extension of these techniques to group III-Nitride systems is important, as their large gap enables high-power/high-temperature applications and high repetition rate SPSs covering a broad spectral range. Also, the high sound velocities and the stronger electromechanical coupling coefficients of nitrides, as compared to (Al,Ga)As materials[16], allow for

---

high frequency applications. While experiments on modulation of the electronic properties of GaN films[17] as well as transport of charge carriers in GaN nanowires[5] by SAWs have been reported, studies on acoustically driven modulation of the optical emission of nitride-based QDs have not yet been demonstrated.

In this letter we use a SAW to periodically modulate the emission wavelength of individual InGaN QDs immersed in GaN nanowires. The dynamic strain field of the SAW transferred to the QDs results in an alternating shift of the QD transition energy at the acoustic frequency of ~330 MHz within a bandwidth up to ~1.5 meV. The small difference in the modulation amplitudes of the exciton (X) and biexciton (XX) QD lines indicate the influences of the SAW fields on the biexciton binding energy. The energy splitting of the X and XX emission scales linearly with the acoustic amplitude, showing that the main effect on the QD electronic structure is due to the strain field of the SAW, with a possible contributions of the SAW piezoelectric field along the nanowire c-axis. The nitride-based dot-in-a-nanowire heterostructures presented here are efficient SPS[18] and can be precisely arranged in periodic two-dimensional arrays[19]. Thus, the present results are an important step towards the development of single photon sources working at high temperature[20], high repetition rates and broad spectral range[18], with simultaneous spatial and time control.

The InGaN/GaN nanowire heterostructures were fabricated by plasma-assisted molecular beam epitaxy (PA-MBE) on (0001) GaN-on-sapphire templates[21]. The nanowires have a typical height of ~500 nm and a diameter of ~200 nm. They exhibit hexagonal cross section with lateral facets defined by non-polar m-planes and a pyramidal top profile formed by six semi-polar r-facets. This profile determines the shape of the InGaN nano-disks embedded inside the GaN nanowire tips[18,21]. The transmission electron microscope (TEM) micrograph of an individual nanowire in Fig. 1(a) reveals the presence of two InGaN sections: a thicker one (~30 nm) formed on the polar facet close to the nanowire top and a narrower (~20 nm thick) one on semi-polar side facets. As detailed in our previous work[18], the QDs under study are formed by fluctuations of the indium content in the topmost InGaN region.

For SAW experiments, the nanowire heterostructures were mechanically transferred onto a SAW delay line consisting of two interdigitated transducers (IDTs) lithographically defined on the surface of a 128° Y-cut lithium niobate (LiNbO$_3$) substrate. A schematic of the device is shown in Fig. 1 (b). We employed floating electrode unidirectional IDTs[22] with a length of ~700 μm and an aperture of ~400 μm (approximately equal to the IDT finger length) designed to generate SAWs with an acoustic wavelength of $\lambda_{SAW} = 11.67$ μm, corresponding to a SAW frequency and period of $f_{SAW} = 338$ MHz and $T_{SAW} = 2.96$ ns, respectively, at the measurement temperature $T = 10\ K$. Note that the exciton decay time in these nanowire-QDs is ~1.3 ns[18], i.e. comparable to $T_{SAW}/2$. The amplitude of the SAW will be specified in terms of the nominal radio-frequency (rf) power ($P_{RF}$) applied to the IDT, without correction for the rf coupling losses. By measuring the rf reflection and transmission spectra using a network analyzer, we determine that only about 30% of the input electrical power applied to the IDT turns into acoustic power. The use of a highly piezo-electric LiNbO$_3$ crystal provides strong strain and electric fields of the propagating SAW, which extend to the optically active nanowire heterostructures deposited on the surface[5,6,23,24].

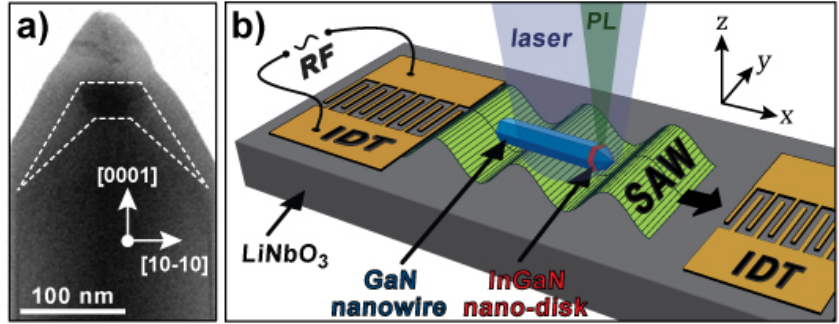

FIG. 1. (a) TEM micrograph of a single nanowire revealing the shape and size of the InGaN nano-disk (marked with dashed lines) embedded inside the nanowire apex[18]. (b) Schematics of the studied device consisting of a LiNbO$_3$ SAW-chip equipped with IDTs and InGaN/GaN nanowire heterostructures dispersed on the SAW propagation path. The Rayleigh-type SAWs are electrically generated by applying an rf voltage to one of the IDTs. The PL is excited under direct illumination by a focused laser beam. The nanowire orientation in (b) is arbitrary and its length is not in scale.

The micro-photoluminescence (μ-PL) experiments were performed at 10 K on a sample mounted in a cold-finger liquid helium flow cryostat equipped with rf connections for the excitation of the IDTs. A continuous wave helium-cadmium laser operating at $\lambda_{exc}$ = 442 nm was used for PL excitation. The laser spot was focused by a 100× microscope objective (NA = 0.73) to a < 1.5 μm diameter spot. The emitted PL was collected by the same objective and dispersed by a single grating monochromator (with an overall spectral resolution of ~350 μeV) equipped with a liquid N$_2$ cooled Si-CCD camera for time-integrated signal detection. For polarization analysis a rotatable half-wave retardation plate and a fixed linear polarizer were placed in front of the monochromator entrance slit, with $0^0$ denoting the SAW propagation direction ($x$-axis in Fig. 1(b)).

Because the excitation energy ($E_{exc}$ = 2.805 eV) is lower than the radiative transitions related to the GaN nanowire[25] and the InGaN region formed on semi-polar side facets[18], carriers are only generated in the apex of the InGaN disk. A typical low-temperature μ-PL spectrum recorded at laser excitation power $P_{exc}$ = 6 μW from dispersed nanowires consists of a series of sharp QD lines on top of a broader background emission. The sharp PL features are attributed to the recombination of photo-generated carriers in QD-like localizing potentials in the topmost InGaN regions[18]. The background emission is due to regions in the InGaN apex without QD-like confinement, as well as different nanowires simultaneously probed by the laser spot. For the experiments discussed in this paper, we focus on a group of QD lines depicted in Fig. 2(a). In the absence of a SAW (upper trace), we observe two narrow (FWHM < 500 μeV) PL peaks appearing at ~2.564 and ~2.587 eV corresponding to the recombination of the exciton (X) and the biexciton (XX) of the same nanowire-QD. This is corroborated by polarization-resolved PL measurements (Fig. 2(b)), which show collinear polarization of the two emission lines with a similar polarization degree (~64%). The linearly polarized emission is due to the valence band mixing induced by the in-plane anisotropy of the QD confinement potential[26]. The peak at 2.585 eV (marked with asterisk in Fig. 2(b)) originates from a different QD, as: *(i)* it does not couple to the SAW field (see below) and *(ii)* it exhibits different polarization direction and ratio (~80%) (cf. Fig. 2(b)). The latter is consistent with the random asymetry of our nanowires QDs formed by InGaN alloy fluctuations[27].

The X and XX assignment is attested by measuring their integrated PL intensities as a function of the optical pump power (cf. Fig. 2(c)). Below the saturation limit of ~6 µW, the 2.564-eV (2.587-eV) emission line develops linear (quadratic) power dependence, expected for the ground-state neutral exciton (biexciton). The energy difference between the X and XX peaks gives a biexciton binding energy $E_{XX}^b = E_X - E_{XX} = -22.9$ meV. The negative binding energy, commonly observed in III-Nitride QDs with inherently large built-in electric fields, is indicative of strong confinement[27,28].

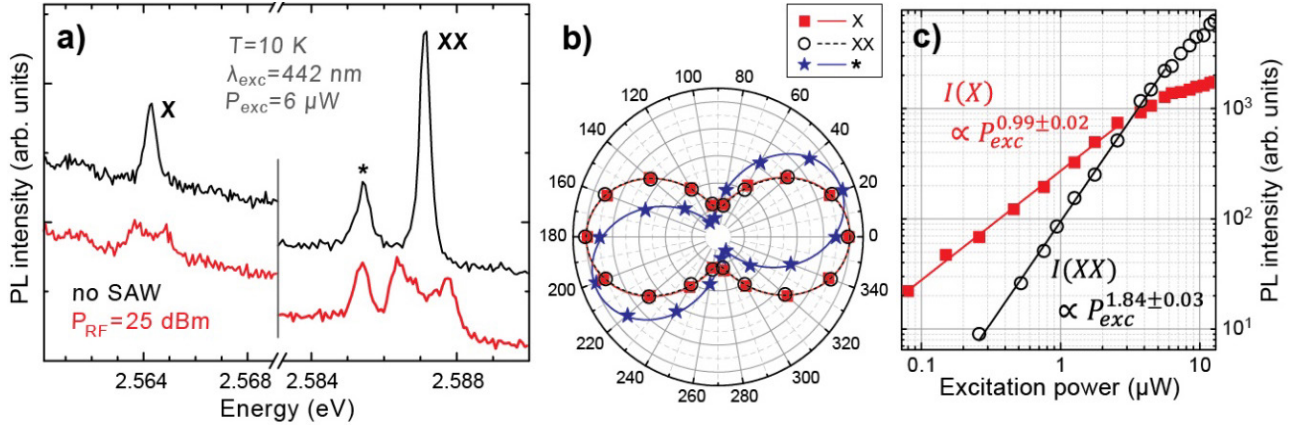

FIG. 2. (a) Low-temperature µ-PL spectrum from a nanowire-QD showing a change in the exciton (X) and biexciton (XX) line structure under the influence of a SAW: upper (lower) trace is recorded in the absence (presence, $P_{RF} = 25$ dBm) of a SAW. The peak marked with an asterisk corresponds to a different nanowire-QD. (b) Polar plots of the integrated peak intensities (with no SAW present) and the corresponding cosine square fits of the PL lines in (a) marked with X (filled squares), XX (open circles) and an asterisk (filled stars). (c) Excitation power dependence of the X (filled squares) and XX (open circles) emission lines in the absence of a SAW. Solid lines are fits to the experimental data.

When the SAW is applied (lower trace in Fig. 2(a)), the emission spectrum undergoes an apparent splitting of the X and XX emission lines into two components. Instead, the PL peak at 2.585 eV remains practically unchanged under the SAW, indicating that it originates from a different nanowire heterostructure within the laser spot, which is weakly affected by the SAW. Actually, the fraction of dispersed nanowires which couple efficiently to the SAW is low (a few percent), probably due to their random orientation resulting in random mechanical contact between the nanowire top (containing the QD) and the underlying substrate. The evolution of the X and XX transition energies with increasing SAW amplitude (which is directly proportional to the square root of the rf power, $\sqrt{P_{RF}}$) is summarized in Figs. 3(a) and 3(b), respectively. In order to discriminate SAW effects from those induced by heating of the sample at high $P_{RF}$, the laser excitation and the rf signal were chopped at the same frequency of 250 Hz[17]. By recording, for each acoustic power, the PL spectrum with the optical and rf excitations in-phase (SAW plus thermal effects) and out-of-phase (only thermal effects), we detected no significant rf-induced thermal contribution on the PL peak position and intensity. The observed changes of the X and XX emission energies can, therefore, be solely attributed to the dynamic modulation of the QD optical transitions induced by the strain and piezoelectric fields accompanying the SAW in LiNbO$_3$ substrate, as reported for (In,Ga)As QDs in Refs.[11,24]. The strain field and its corresponding hydrostatic pressure create alternating regions of maximum compression and tension separated by $\lambda_{SAW}/2$. This, in turn, induces a deformation potential modulation of the QD energy levels, causing a periodic shift of the transition energies with respect to their equilibrium energy in the absence of a SAW. The energy shifts induced by the SAW strain should increase linearly with the acoustic amplitude[11,24] and are expected to be a dominant tuning mechanism at low SAW intensities.

Spectral shifts associated with quantum confined Stark effect (QCSE) governed by the oscillating piezo-electric field of the SAW exhibit in general a quadratic dependence on SAW amplitude[24], and are likely to contribute mostly at large SAW amplitudes. However, in polar c-plane III-Nitride QDs with large polarization-related fields (typically of several MV/cm), electric fields applied parallel[29,30] to the QD polar axis give rise to "linear" Stark effect. In our experiment, according to calculations based on the method described in Ref.[31], a SAW excited at $P_{RF} = 25\ dBm$ produces a dominant strain component (0.05%) along the wave direction (x-axis in Fig. 1(b)). At times during the acoustic cycle corresponding to maximum compressive or tensile strain, only the transverse component (of the order of few kV/cm) of the oscillating SAW piezoelectric field (i.e. along z-axis in Fig. 1(b)) is present. To try to distinguish between the effects of strain and piezoelectric field during the SAW half-cycles corresponding to maximum tension and compression, the emission spectrum in Fig. 2(a) was recorded for different laser excitation powers in the absence and presence of a SAW (Fig. 3(c)). When no SAW is applied the X emission energy undergoes an apparent blueshift (up to 300 μeV) with increasing pump power (empty squares), indicating partial screening of the large internal electric field by carriers photo-generated in the InGaN regions. For nanowire-QDs with the c-axis parallel to the SAW piezoelectric field, the energy splitting induced by the acousto-electric effect is also expected to change at high optical pumping. No change in splitting is observed in our case even at high excitation intensities (well above the X and XX saturation limit), as shown for the X line in Fig. 3(c) (filled squares). This suggests that the electric field parallel the c-axis has a minor influence on the observed energy splitting of the X and XX peaks. Considering that the excitonic transitions in our nanowire-QDs are polarized in the growth plane (i.e. perpendicular to the nanowire axis)[18], the polarization-resolved measurements in Fig. 2(b) imply that the nanowire under study is perpendicular to the SAW propagation direction. However, its exact orientation is difficult to determine due to the presence of serval nanowires within the excitation spot. We can exclude, in principle, any effect of the SAW piezoelectric field perpendicular to the QD polar axis, if we assume QD symmetry around the c-axis. In such case, the effect of this field should not depend on its direction (i.e. positive or negative field)[32], which is contrary to the observed symmetric splitting of the X and XX peaks around their equilibrium energies when no SAW is applied (Fig 3(a) and (b)).

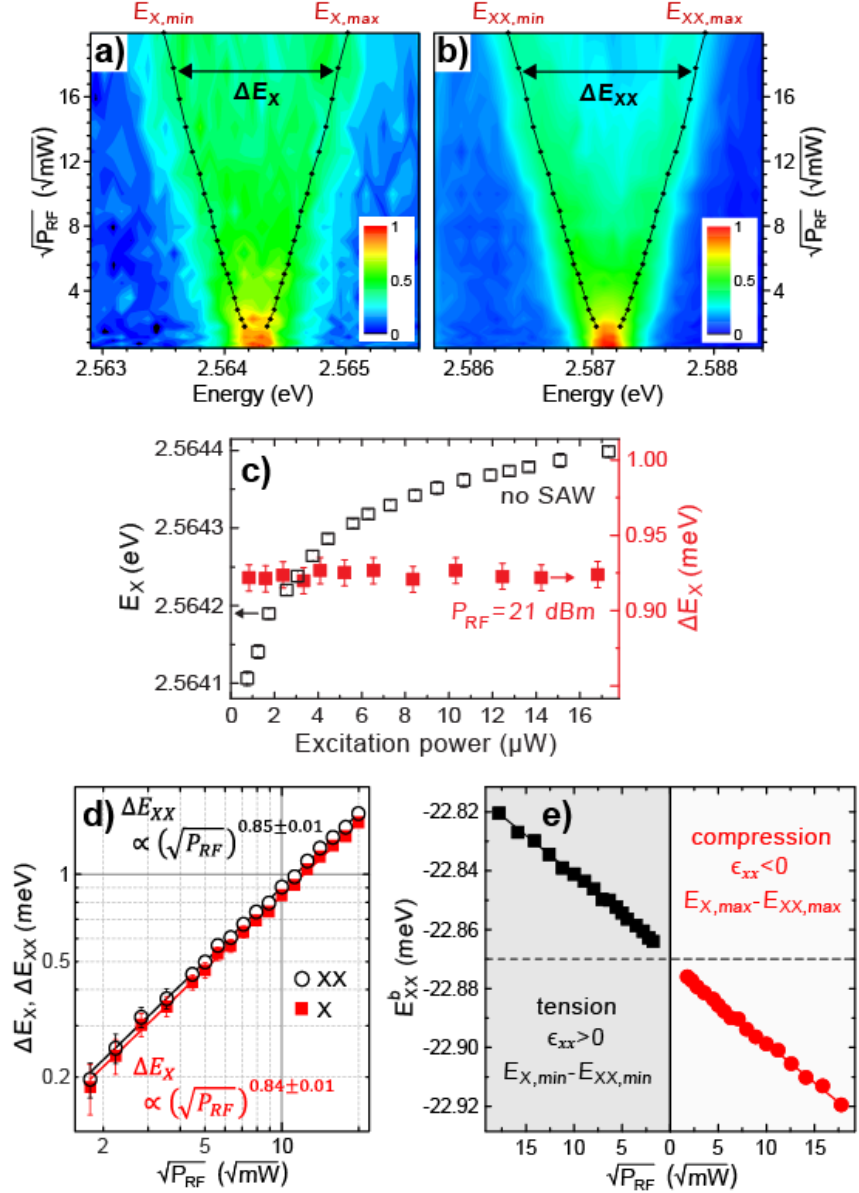

FIG. 3. False-color plot of X (a) and XX (b) emission lines as a function of the SAW amplitude ($\propto \sqrt{P_{RF}}$) showing a pronounced splitting of the excitonic transitions into a doublet consisting of lowest ($E_{X(XX),min}$) and highest ($E_{X(XX),max}$) energy values separated by $\Delta E_{X(XX)}$. The $E_{X(XX),min}$ and $E_{X(XX),max}$ values for different acoustic powers are indicated by dots. (c) Excitation

power dependence of the exciton energy (open squares) and SAW-induced splitting (filled squares). (d) Energy splitting of the X (squares) and XX (open circles) emission lines extracted from (a) and (b), respectively, as a function of $\sqrt{P_{RF}}$. Lines are power-law fits to the experimental data indicative of deformation potential coupling. (e) Acousto-mechanically induced variation of the XX binding energy ($E_{XX}^b$). Solid lines are guides to the eye, indicating a decrease (increase) of $E_{XX}^b$ under compressive (tensile) strain. The dashed line denotes unperturbed $E_{XX}^b$ value in the absence of a SAW.

The evolution of the spectra of Figs. 3(a) and 3(b) can thus be associated with the acousto-mechanical modulation of the QD band structure together with a possible contribution of the piezoelectric field in the linear Stark effect regime. However, this contribution seems to be small, as shown by the lack of screening mentioned above. The energy separation between split doublet lines ($\Delta E_{X(XX)} = E_{X(XX),max} - E_{X(XX),min}$) reflects the difference between the transition energies corresponding to maximum compressive ($E_{X(XX),max}$) and tensile ($E_{X(XX),min}$) strain. The values of $\Delta E_X$ (full squares) and $\Delta E_{XX}$ (open circles) extracted from Figs. 3(a) and 3(b), respectively, are plotted in Fig. 3(d) as a function of the SAW amplitude. The magnitude of $\Delta E_X$ ($\Delta E_{XX}$) continuously increases with increasing $P_{RF}$, reaching a maximum of $\Delta E_X = 1.51$ meV ($\Delta E_{XX} = 1.62$ meV) for the highest SAW intensity at $P_{RF} = 25$ dBm. From this $\Delta E_X$ value, using deformation potentials for InGaN[33] and neglecting the contribution form the SAW piezoelectric field, we estimate an approximate strain value in the QD of the order of 0.04%. This value is comparable to the calculated 0.05% strain component at the LiNbO$_3$ surface along the SAW propagation, thus indicating high mechanical coupling of the present nanowire to the SAW. Contrary to the ideal linear trend expected for deformation potential coupling, the sub-linear dependence of $\Delta E_X$ and $\Delta E_{XX}$ on the acoustic amplitude (solid lines in Fig. 3(d)) can be ascribed to both quantum confinement-induced (also responsible for linearly polarized X and XX emission) and SAW induced valence band mixing. In analogy to GaAs/AlGaAs QWs[2], for mixed hole states, the modulation of the valence-band-edge deviates from a sinusoidal shape resulting in the observed behavior.

The small difference between $\Delta E_X$ and $\Delta E_{XX}$ in Fig. 3(d) reflects the variation of the biexciton binding energy $E_{XX}^b$ upon acoustic excitation. This becomes evident in Fig. 3(e) which displays the $\sqrt{P_{RF}}$-dependence of $E_{XX}^b$ under maximum compressive (circles) and tensile (squares) strain. The modulation span of the biexciton binding energy over the entire range of $P_{RF}$ reaches 100 µeV (Fig. 3(e)). The biexciton binding energy is roughly given by $E_{XX}^b = 2J_{eh} - J_{ee} - J_{hh}$, where $J_{ij}$ stands for the direct Coulomb interaction between carriers $i$ and $j$. Compression (tension) along the QD polar axis would result in an increase (decrease) of the electron-hole attraction $J_{eh}$ thus increasing (decreasing) $E_{XX}^b$[27,34,35]. The opposite holds for strain perpendicular to the polar axis, which would mainly affect the repulsive electron-electron ($J_{ee}$) and hole-hole ($J_{hh}$) interactions. Consequently, changes in $E_{XX}^b$ with increasing SAW intensity (Fig. 3(e)) are consistent with the effect of lateral size variation of the confinement potential on the biexciton binding energy, as reported for III-Nitride QDs[27,34,35]. The observed trend can thus be explained by the effect of the SAW strain field and some possible contribution of the SAW-induced electric field parallel to the c-axis on the QD confining potential[30], affecting the Coulomb interactions within the four-particle biexciton complex.

In summary, we have demonstrated the SAW-driven modulation of the optical emission of a single GaN/InGaN nanowire-QD. We show that the acousto-mechanical coupling shifts the QD energy levels giving rise to a characteristic splitting (up to 1.5 meV) of the excitonic transition energies. The SAW fields also induce a monotonic change of the biexciton binding energy without any appreciable degradation of the emission, thus providing a reliable tool for in situ control of the exciton-biexciton system. By collecting photons in a narrow energy range, the dynamic spectral tuning reported here can be readily used to control the QD emission times and obtain triggered single photon sources operating at acoustic frequencies without the need for a pulsed laser, as reported for III-As QDs in Ref.[8].

We thank A. Hernández-Mínguez and P. V. Santos from the Paul-Drude-Institute, Berlin (Germany) for preparation of the SAW delay lines. E.C. (S.L.) acknowledges the Spanish MINECO FPI (RyC-2011-09528) grant. Financial support from the Spanish MINECO (contracts MAT2011-22997 and MAT2014-53119-C2-1-R) and the CAM (contract S2009/ESP-1503) is gratefully acknowledged.